\documentclass[sigconf]{acmart}

\usepackage[english]{babel}
\usepackage{blindtext}
\usepackage{graphicx}
\usepackage{subfigure}
\usepackage{threeparttable}
\usepackage[justification=centering]{caption}

\renewcommand\footnotetextcopyrightpermission[1]{} 
\setcopyright{none}

\acmDOI{}

\acmISBN{}

\acmPrice{}

\begin{document}
\title{A Study on the Characteristics of Douyin Short Videos and Implications for Edge Caching}


%
\author{Zhuang Chen}
\affiliation{%
  \institution{Guilin University of Electronic Technology}
  \city{Guilin}
  \state{China}
}
\email{zhuangchenuio@gmail.com}

\author{Qian He}
\affiliation{%
  \institution{Guilin University of Electronic Technology}
  \city{Guilin}
  \state{China}
}
\email{heqian@guet.edu.cn}

\author{Zhifei Mao}
\affiliation{%
  \institution{Norwegian University of Science and Technology (NTNU)}
  \city{Trondheim}
  \country{Norway}}
\email{zhifei.mao@gmail.com}

\author{Hwei-Ming Chung}
\affiliation{%
  \institution{University of Oslo}
  \city{Oslo}
  \country{Norway}
}
\email{hweiming.chung@gmail.com}

\author{Sabita Maharjan}
\affiliation{%
 \institution{University of Oslo, and Simula Research Laboratory}
 \city{Oslo}
 \country{Norway}
 }
 \email{sabita@simula.no}



\begin{abstract}
Douyin, internationally known as TikTok, has become one of
the most successful short-video platforms. To maintain its popularity, Douyin
has to provide better Quality of Experience (QoE) to its growing
user base. Understanding the characteristics of Douyin videos is
thus critical to its service improvement and system design.
In this paper, we present an initial study on the fundamental
characteristics of Douyin videos based on a dataset of over
260 thousand short videos collected across three months. The
characteristics of Douyin videos are found to be significantly
different from traditional online videos, ranging from video
bitrate, size, to popularity. In particular, the distributions
of the bitrate and size of videos follow Weibull distribution.
We further observe that the most popular Douyin videos
follow Zifp's law on video popularity,
but rest of the videos do not. We also investigate the correlation
between popularity metrics used for Douyin videos.
It is found that the correlation between the number of views and the number of likes are strong, while other correlations are relatively low.
Finally, by using a case study, we demonstrate
that the above findings can provide important guidance on designing an efficient edge caching system.
\end{abstract}


\maketitle

\section{Introduction}
The success of online video sharing platforms (e.g., YouTube, Facebook, Instagram, and Snapchat) has been phenomenal. According to Cisco's annual Visual Network Index (VNI) forecast~\cite{cisco}, video accounts for an overwhelming share of total Internet traffic. With the increasing usage of mobile devices, video traffic moves from wired ends (e.g., PCs) to mobile ends (e.g., smartphones). 
With this shift to the mobile Internet, the video sharing industry has been reshaped in recent years. One of the biggest trends is the emergence of short-form video platforms (e.g., Douyin). These platforms have typically a large number of User Generated Content (UGC) with few tens of seconds that is two orders of magnitude shorter than the length of a typical traditional video. 

The success of similar mobile Internet applications like Douyin, Instagram, and YouTube, 
depends on rich video libraries, and even more importantly, for optimally designing the caching system. 
Moreover, short video services with ever-increasing popularity use a large portion of Internet bandwidth. 
Besides, they are time-sensitive, especially for short video platforms like Douyin. 
Douyin has now exceed 150 million active daily users in China, while the average size of a video file uploaded is 1.96 MB~\cite{b}. If every user only uploads a 1.96 MB video every day, the total disk space required to store all the videos is at least about 294 TB. 
Therefore, dynamic and efficient caching is necessary for such platforms.
In addition, bandwidth cost and end-to-end latency are equally important issues for Douyin.
QoE is no doubt the biggest challenge it faces. 
Edge caching~\cite{YanZhang} can not only reduce the usage of backhaul bandwidth and the delay, but also improve energy efficiency, that is vitally important for capacity planning and for QoE enhancement. 

Established in 2016, Douyin, also known as TikTok, has become one of the fastest-growing  mobile Internet applications. Industry insiders estimated that Douyin outstripped YouTube, Facebook, Instagram, and Snapchat in total downloads in September 2018~\cite{2018}, and Sensor Tower estimates that Douyin has surpassed one billion installs on the App Store and Google Play in February 2019~\cite{2019}.
On the other hand, some studies such as YouTube~\cite{Koch:2018:CHC:3204949.3204963} and Twitter~\cite{8353149}, have analyzed different characteristics, for studying caching mechanism.
Different from YouTube and Twitter, Douyin is specially designed to provide short videos for the mobile Internet users.
While the caching mechanism for Douyin is based on the studies of traditional short videos, three distinct features of Douyin call for novel design of the cache for Douyin.
First, the number of Douyin videos is much larger compared to the number of traditional short videos. 
Second, the size of Douyin videos 
is much smaller than conventional short videos (90\% being less than 1.5 MB, while a typical YouTube short video is 25 MB~\cite{6522525}).
Finally, the viewing frequency of the most popular Douyin videos fits the well-known Zipf distribution, which can have an important effect on edge caching for short videos  ~\cite{749260}.
Considering the growing popularity and use of the platform, understanding the characteristics of Douyin videos is therefore of important for designing a dynamic and efficient caching system for the platform.

In this paper, we present an initial study on the fundamental characteristics of Douyin videos. 
We analyze the features of video file and popularity on a dataset of over 26000 short videos collected through a three-month period in early 2018. We show that Douyin video bitrate and size can be modeled as Weibull distribution\cite{scholz2008inference}. 
We also look closely at the popularity metrics of Douyin videos, in terms of number of views, number of likes, number of comments and number of shares.  
The correlation analysis using Pearson coefficient indicates that most of the popularity metrics have minimal correlation, except the number of views and the number of likes.
Nevertheless, by using a case study, it is suggested that our analysis in this paper can serve as guidelines for designing efficient caching system tailored specifically for the unique characteristics of Douyin short videos.

To the best of our knowledge, our work is the first to study Douyin short videos, which not only provides a basis for further exploring and understanding Douyin, but also provides an initial foundation for the design of edge caching systems for the the latest short video platforms.

Our main contributions in this paper can be summarized as: 
\begin{itemize}
  \item We provide the first and extensive characterization of Douyin short videos, based on a real-world dataset from Douyin. 
  \item We examine the popularity distribution of Douyin videos, which will be of special importance to design a popularity-based caching system for Douyin.
  \item We further analyze the relationship between popularity metrics used for Douyin videos. The analysis result suggests that a new computing paradigm for video popularity is indeed needed. In addition, a case study for edge caching is designed based on the above results, which further confirms the applicability of our research. Moreover, the case study suggests that the results of our study can be  highly beneficial to the design and development of an efficient and intelligent caching system for the latest form of mobile social short video media like Douyin.
 \end{itemize}

The remainder of the paper is organized as follows. In Section 2, a brief background on Douyin, and its video dataset is presented. We study the characteristics of Douyin short videos in Section 3. We analyze the popularity of Douyin videos in Section 4. We conduct a case study on edge caching in section 5. Finally, we conclude the paper with an outlook towards future work in Section 6.

\section{Background}
In this section, first we provide a short introduction of Douyin. 
Then the dataset of Douyin short video is described.

\subsection{Douyin}
Douyin is a mobile short video platform with powerful editing capabilities, which enables users to add various types of music and effects on their videos. The length of Douyin videos is restricted to 15 seconds, which makes them more attractive.

The content delivery mechanism of Douyin is decentralized. When receiving the user-uploaded videos, Douyin ranks them and recommends the relevant short videos by analyzing the interests of the user.
Douyin contains a large amount of UGC. For a large amount of short video content, the recommendation mechanism calculates the tag for each video, which is designed to classify videos according to category characteristics. Then, it maps the tag of the video to the users who have the same tag. 

\subsection{Douyin Short Video Metadata}
Our dataset consists of the metadata of short videos. In addition to the history of the short videos uploaded, Douyin also archives users' profiles and their social networks including followed users and their fans.

Douyin assigns a distinct 19-digit  decimal ID for each video. Each video contains the following meta-data: video ID, the time when it was released, bitrate which is the play bitrate of each video, video length, which is the play duration of each video, video file size that is one of the key metrics for caching, verification type that indicates whether the user who uploaded the video, has passed the official certification of Douyin, number of views and number of likes, number of comments and number of shares. The basic parameters of the meta-data is shown in Table 1. 

The data was collected from 1st February, 2018 to 10th May, 2018, including 270 thousand videos from different users. After removing repeated videos, we got 260939 videos. Each entry contains all the meta-data except the video size.
 
\begin{table}[htbp] 
\caption{Meta-Data of A Douyin Video}
\begin{center}
\begin{tabular}{|l|l|}

\hline
Video ID& $6553843141084974340$ \\

\hline
Video Release Time& 	May 10, 2018, 14:58:00\\

\hline
Bitrate	&$1104867$ bps\\ 

\hline
Video Length(Duration)&	$15070$ ms\\

\hline
Video File Size	&$1.98$ MB\\

\hline
Verification Type	 &$1$\\

\hline
Number of Views	&$1564$\\

\hline
Number of Likes	&$12$\\

\hline
Number of Comments	&$3$\\

\hline
Number of Shares	&$1$\\
\hline

\end{tabular}
\label{tab1}
\end{center}
\end{table}

\begin{table}[t]
\begin{center}
\begin{threeparttable}
\caption{Statistics of Video Length, Bitrate and Size} \label{tab:cap}
\footnotesize  
\begin{tabular}{|l|r|r|r|r|r|}
  \hline
   & Min & Max & Mean & Median & Std.Dev.
  \\
  \hline
   length(s) & 4 & 73 & 13.1 & 14 & 3.9 \\
   bitrate(kbps) & 0 & 4719 & 1271.6 & 1205 & 691.3 \\
   size(MB) & 0 & 24.5 & 1.96 & 1.8 & 1.2 \\
  \hline
\end{tabular}
\small Note: Std.Dev. is short for Standard Deviation.
\end{threeparttable}
\end{center}
\end{table}

\section{CHARACTERISTICS OF Douyin SHORT VIDEO}
In this section, we characterize the Douyin video files, in terms of video length, video bitrate and video size. 
The characteristics of Douyin videos can be classified into two types: time-invariant and time-variant. Some characteristics are static, such as video length, video file size, and video published time, while others are dynamic, e.g., number of views, number of likes, number of comments and number of shares. However, the information is static within each time slot. The following characteristics are studied: video length, video file size, number of views, number of likes, number of comments, and number of shares. In addition, we also investigate the relationships among them.

\subsection{Video Length}
The length of Douyin video is one of the most significant differences compared to traditional videos, which narmally last 0.5-2.5 hours (e.g., YouTube~\cite{6522525}), Douyin mainly provides short musical videos, 95\% of the video length in our entire datasets are within 15 seconds imposed by Douyin on regular user uploads. However, we do find videos that exceeded this limitation, because Douyin officially permits a small group of authorized users to upload videos longer than 15 seconds.

Fig. 1 shows the probability density function (PDF) and cumulative distribution function (CDF) of the Douyin videos' length within 70 seconds, which exhibits two peaks. The highest peak is between 14 and 16 seconds, which accounts for about 65\% of the videos. In addition, it is clear that the videos of 15 seconds is the most popular among users. The second peak is between 9 and 11 seconds, accounting for about 27\% of the total.

\begin{figure}[htbp]
\centerline 
{\includegraphics[width=3.5in]{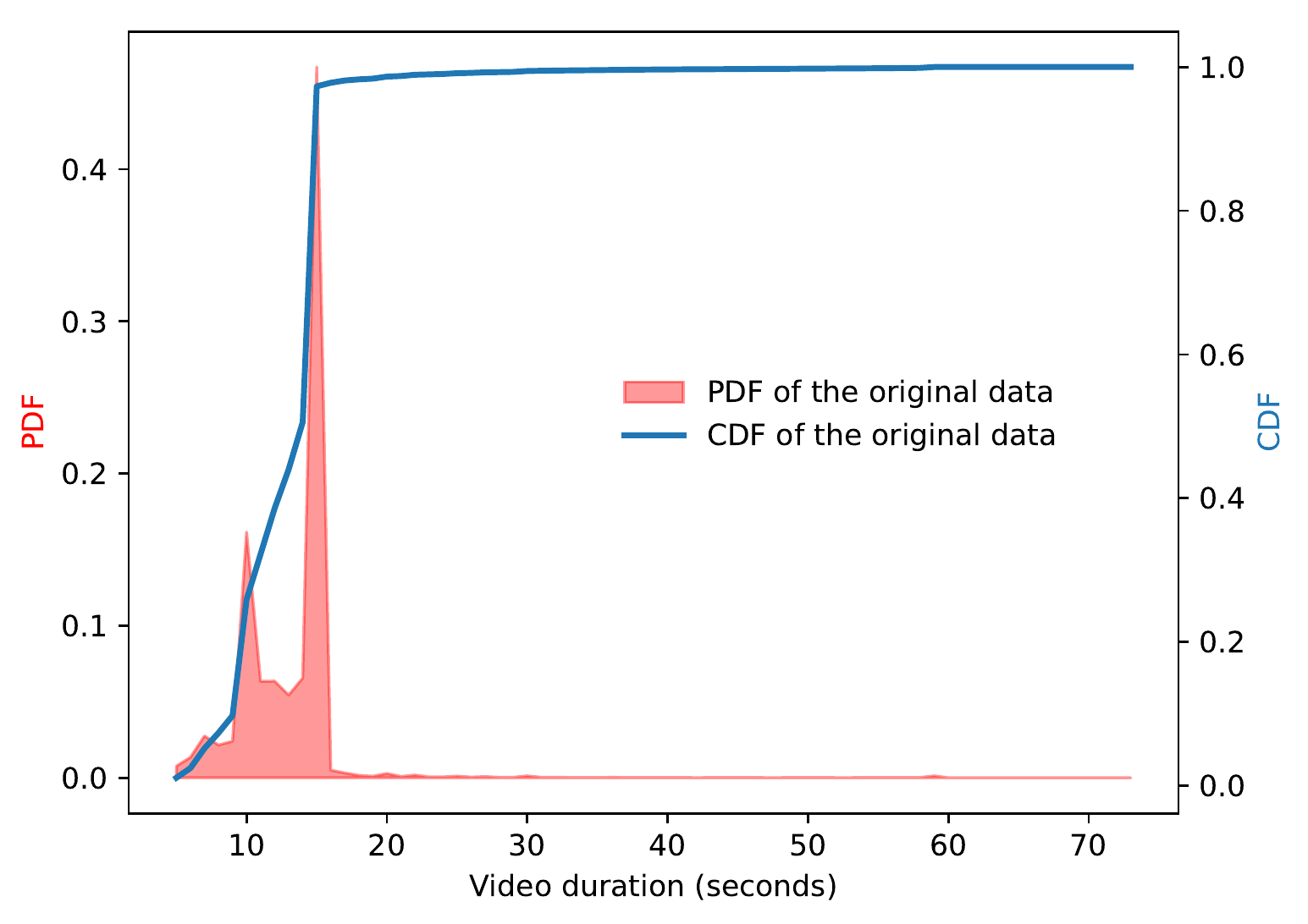}} 
\caption{Distribution of Douyin Video Lengths.}
\label{Figure 1}
\end{figure}

\begin{figure}[htbp]
\centerline
{\includegraphics[width=3.5in]{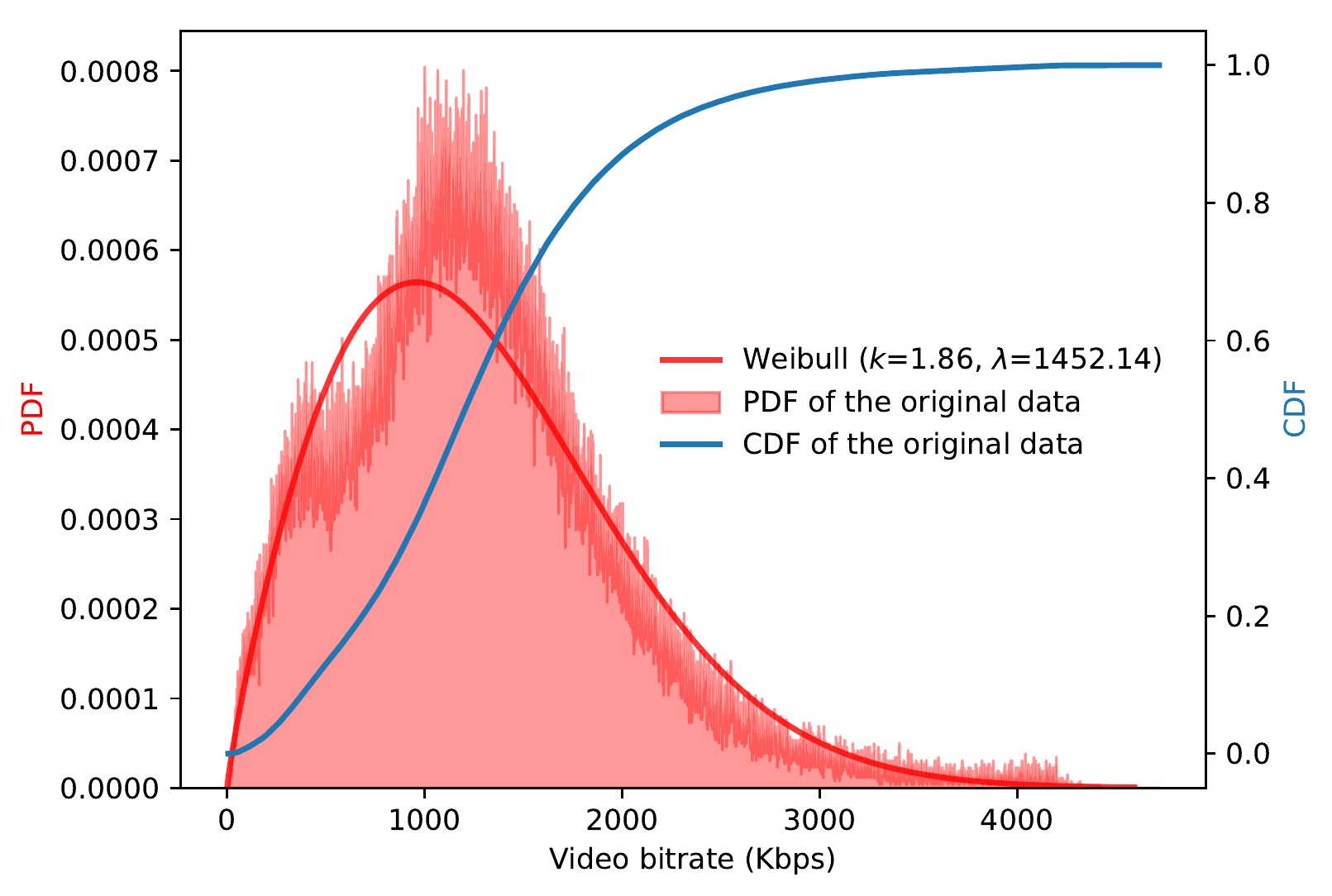}} 
\caption{Distribution of Douyin Video Bitrates.}
\label{Figure 2}
\end{figure}

\begin{figure}[htbp]
\centerline
{\includegraphics[width=3.5in]{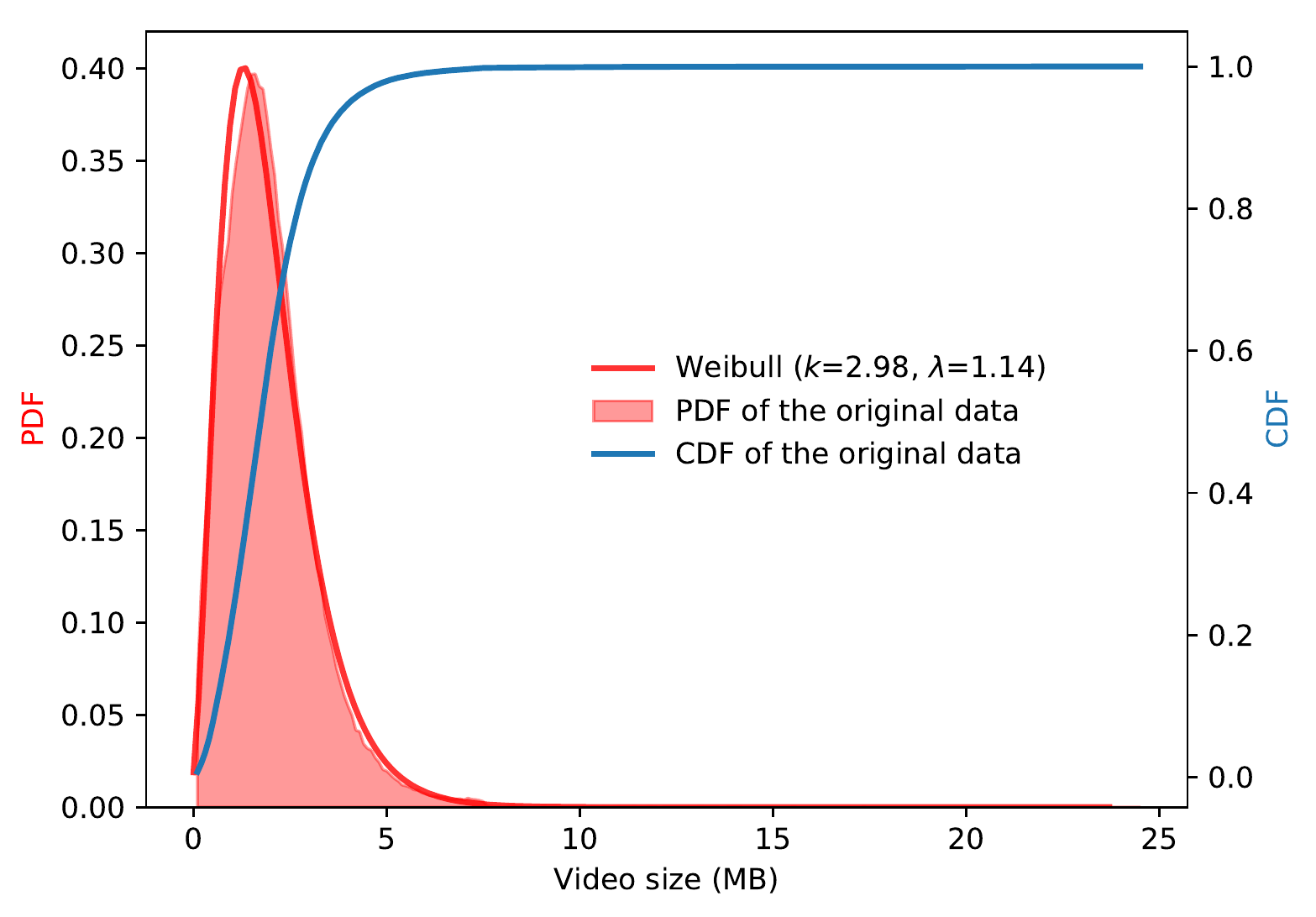}} 
\caption{Distribution of Douyin Video File Size.}
\label{Figure 3}
\end{figure}

\subsection{Bitrate}
The bitrate of a video is an indicator of its playback quality. 
Low bitrate degrades user's QoE~\cite{Gill:2007:YTC:1298306.1298310}, leading to decline in the popularity of Douyin over time.
We observe that Weibull distribution fits the skewed curve of the bitrate of Douyin videos. 
This insight can be useful in designing an efficient caching system based on adaptive bitrate~\cite{8411495}.

Fig. 2 shows there are two peaks of bitrate among the videos viewed. One peak is around 1130 kbps, and the other one is around 410 kbps. Only about 2.7\% of the videos are encoded at lower bitrates below 200 kbps. Similarly, approximately 1.8\% of the videos are encoded at a higher bitrate than 3000 kbps. This implies that it does not follow the well-known Zipf distribution. 
Compared to conventional videos, Douyin videos have higher bitrates i.e., 74.1\% of the videos have bit rates between 500 Kbps and 3000 Kbps, which is possibly due to the development of network communication technology and the enhancement of the function of the device chip. 
In the near future, the widespread commercialization of 5G communication technology will further increase the range of bitrate for multimedia video. 

\subsection{Video File Size}
The video file size information is not available for Douyin videos. However, we can calculate the video file size from video length (duration) and its bitrate. The size of each video file can be calculated as follows
\begin{equation}
Size = bitrate \times length                     
\end{equation} 

As illustrated in Fig. 3, we plot the PDF and CDF of video file sizes, and find that the distribution of video file sizes is different from the distribution of video lengths, even if there is a direct relationship between them.  In the collected dataset, 97.8\% of the videos are smaller than 5 MB. Based on Table 2, an average video file size is 1.96 MB, which is smaller than that of the YouTube videos (7.6 MB)~\cite{6522525}. However, considering that there are 150 million active users in a day, if every user uploads a 1.96 MB video, the total disk space required to store all the videos is at least 294 TB everyday. Therefore, efficient caching is essential. 
We also list the statistics of video length, bitrate, and size in Table 2.

\begin{figure}[htbp]
\centering
\subfigure[PMF of video views.]{
\begin{minipage}[t]{0.48\linewidth}
\centering
\includegraphics[width=1.5in]{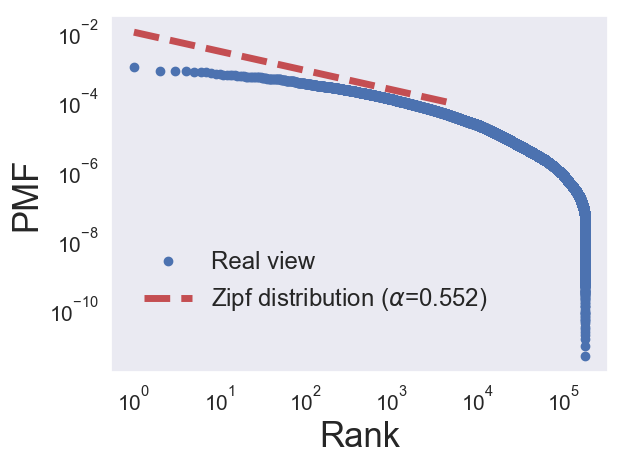}
\end{minipage}%
}%
\subfigure[PMF of video likes.]{
\begin{minipage}[t]{0.48\linewidth}
\centering
\includegraphics[width=1.5in]{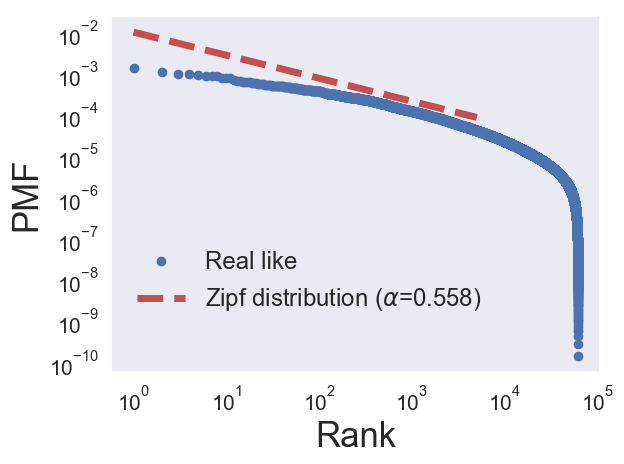}
\end{minipage}%
}%

\subfigure[PMF of video comments.]{
\begin{minipage}[t]{0.48\linewidth}
\centering
\includegraphics[width=1.5in]{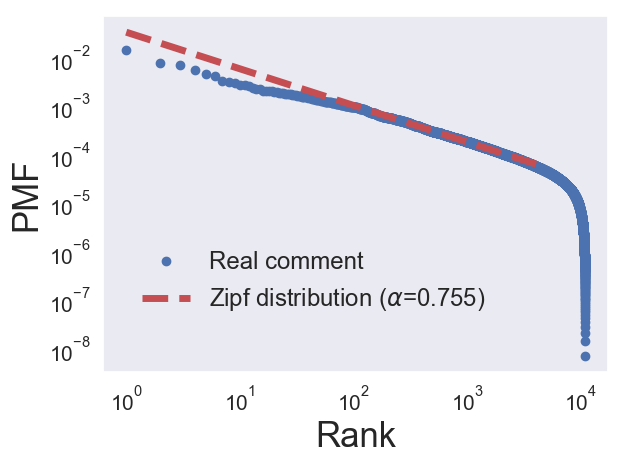}
\end{minipage}
}%
\subfigure[PMF of video shares.]{
\begin{minipage}[t]{0.48\linewidth}
\centering
\includegraphics[width=1.5in]{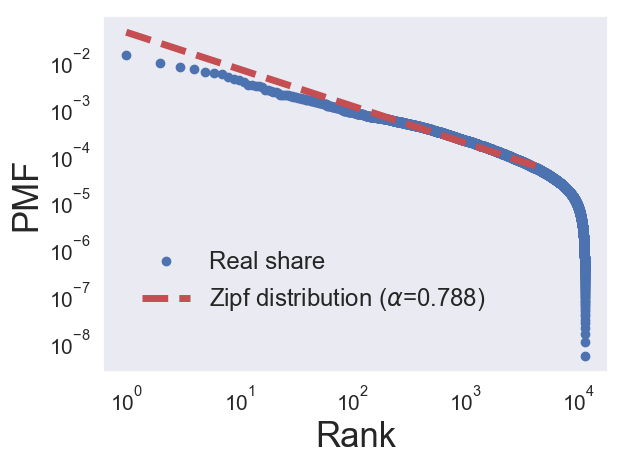}
\end{minipage}
}%
\centering
\caption{ Distributions of popularity metrics.}
\end{figure}

\begin{figure}[htbp]
\centering
\subfigure[CDF of video views.]{
\begin{minipage}[t]{0.48\linewidth}
\centering
\includegraphics[width=1.5in]{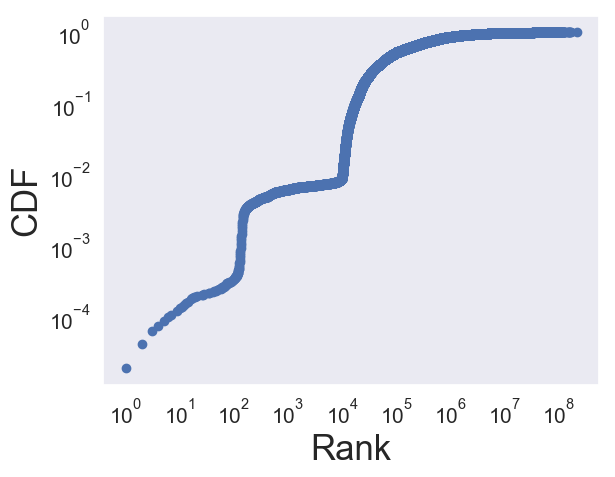}
\end{minipage}%
}%
\subfigure[CDF of video likes.]{
\begin{minipage}[t]{0.48\linewidth}
\centering
\includegraphics[width=1.5in]{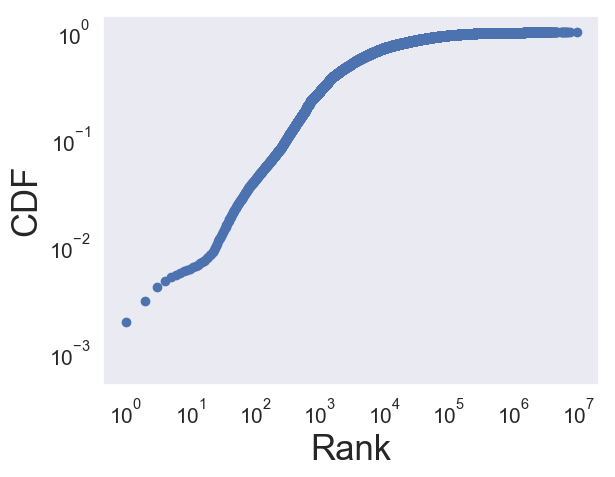}
\end{minipage}%
}%

\subfigure[CDF of video comments.]{
\begin{minipage}[t]{0.48\linewidth}
\centering
\includegraphics[width=1.5in]{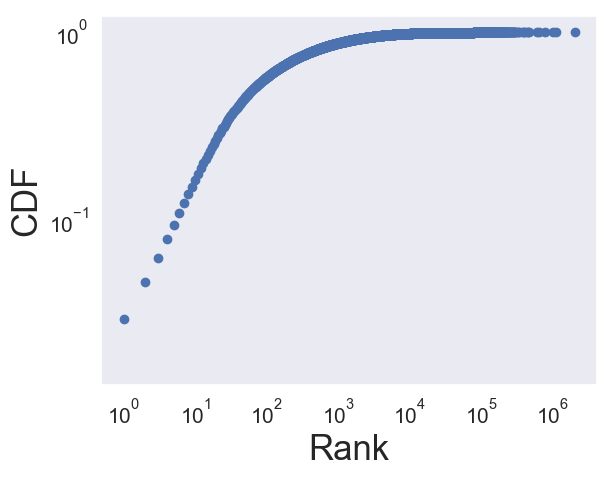}
\end{minipage}
}%
\subfigure[CDF of video shares.]{
\begin{minipage}[t]{0.48\linewidth}
\centering
\includegraphics[width=1.5in]{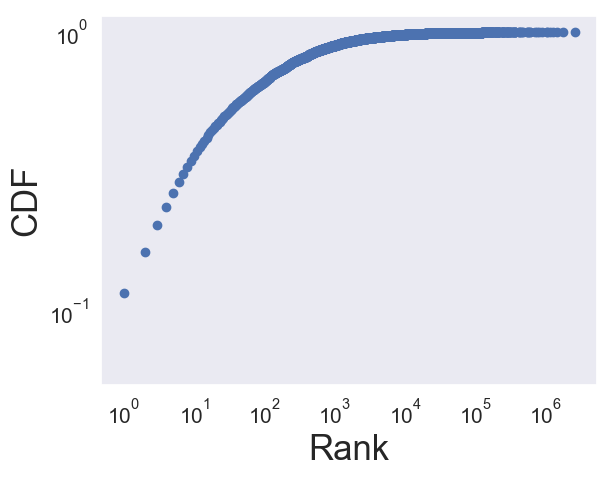}
\end{minipage}
}%
\centering
\caption{ CDF of popularity metrics.}
\end{figure}

\section{ Video Popularity}
Video popularity plays an important role in the design of recommendation system and cache mechanism. Popular videos are likely to be recommended to the users and cached at the edge servers close to the users in order to reduce delay. In this section, we analyze the popularity of Douyin short videos. 

\subsection{Distribution}
There are four popularity indicators available for each Douyin video: number of views, number of likes, number of comments, and number of shares.
We rank all the videos in terms of the above popularity indicators, normalize the popularity values and plot on a log scale in Fig. 4, it can be easily seen from Fig. 4 that the distributions do not follow Zipf's law which appears approximately linear on log-log plot. However, we conformed that the distribution of 5000 most popular videos follow the Zipf's law (In the next section, we will explain the rationale to approximate the distributions of only the most popular videos and how such approximation can be leveraged in designing the caching system.). Mathematically, Zipf's law can be defined by
\begin{equation}
\begin{aligned}
&p_n \sim n^{-\alpha}\\       
\end{aligned}                         
\end{equation}
meaning that the popularity of the $n$th most popular video $p_n$ is ${\frac{1}{n^{\alpha}}}$ the popularity of the most popular video, where $\alpha$ is a constant parameter. As seen in Fig. 5 (a), by doing least squares polynomial fitting, we found that Zipf's law with $\alpha = 0.552$ fits very well to our empirical observations. The normalized number of views of the three most popular videos are $p_1=0.022$, $p_2=0.017$, $p_3=0.016$, respectively. It is easy to see that $p_2\sim \frac{1}{2^{0.552}}*p_1$ and $p_3\sim \frac{1}{3^{0.552}}*p_1$.

From above analysis, one may conclude that the most popular videos take away most of the views as well as likes, comments and shares. 
This is also verified by Fig. 5.
For example, $18.6\%$ most popular videos unproportionally account for $80.5\%$ of all the views. It means that most of the videos get very few views compared to the popular ones.

\subsection{Correlation}
Is a video that is popular in terms of number of views also popular in terms of number of likes, number of comments, or number of shares? To answer this question, we use the Pearson correlation coefficient to study the correlation of the four popularity indicators. The Pearson correlation coefficient between two variables $X$ and $Y$ is given by 
\begin{equation}
\begin{aligned}
\rho \left (X,Y\right ) =  \frac{E[XY] - E[X]E[Y]} {\sqrt{E[X^2] - [E[X]]^2} \sqrt{E[Y^2] - [E[Y]]^2}}         
\end{aligned}              
\end{equation} 
The Pearson correlation coefficient is unitless and ranges from -1 to 1. Higher coefficient indicates higher correlation.

Fig. 6 shows the correlation coefficients between each two of the four popularity indicators. We can see that all correlation coefficients are positive. In particular, the number of views and the number of likes have a very high correlation coefficient which is 0.91, meaning that a video which is popular in terms of number of views is very likely to be popular in terms of number of likes and vice versa. In comparison, other coefficients are relatively low. Especially, the correlation between the number of shares and the number of comments is very low. 

\begin{figure}[htbp]
\centerline 
{\includegraphics[width=3.5in]{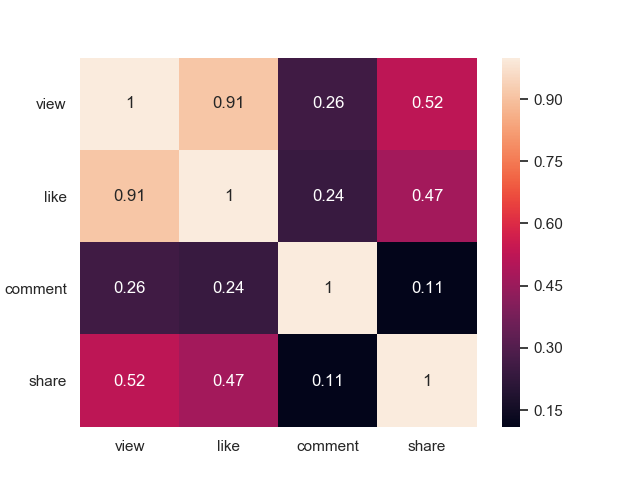}} 
\caption{Correlation coefficient of popularity metrics.}
\label{Figure 6}
\end{figure}

\section{Edge Caching: A Case Study}
The above study on Douyin has important implications on improving Douyin's services. As the user base of Douyin is expanding rapidly, the content servers are facing challenge to answer users' requests timely. In the 5G era, it is possible to cache contents at the edge servers which make services closer to the end users. In this section, we show the feasibility and benefits of caching popular short videos at the edge. 

\begin{figure}[htbp]
\centerline 
{\includegraphics[width=3in]{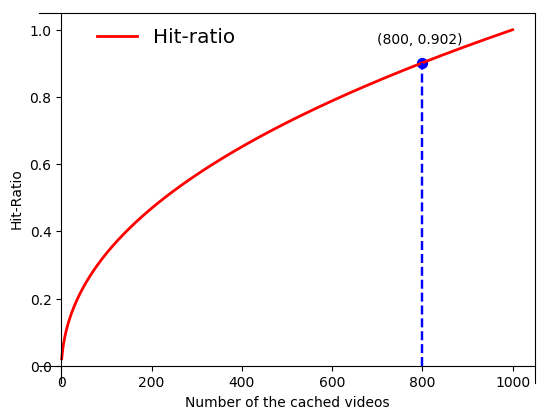}} 
\caption{Hit-ratio as a function of the number of cached videos.}
\label{Figure 7}
\end{figure}

Assume that the probability of a video with rank $\kappa$ is watched in the next time slot $\tau$ is $ p^\tau_\kappa = \frac{V^\tau_\kappa}{V^\tau} $, where $V^\tau_\kappa$ is the number of views on a video with rank $ \kappa $ and $ V^\tau $ is the total number of views of all videos. Fig. 5 (a) has already shown that the number of views of the most popular videos follows Zipf's law with the characteristic exponent $ \alpha = 0.552 $. Then, $ V^\tau_\kappa = V^\tau(\frac{1/\kappa^{0.552}}{\sum_{i=1}^{N_\tau}1/i^{0.552}}) $ where $ N_\tau $ is the number of videos and $ p^\tau_\kappa = \frac{1/\kappa^{0.552}}{\sum_{i=1}^{N_\tau}1/i^{0.552}} $ is the Zipf's law function. Given 1000 Douyin videos, the hit-ratio of caching the most popular video is as high as 2.1\% ( $ p^\tau_{1} = \frac{1/1^{0.552}}{\sum_{i=1}^{1000}1/i^{0.552}} = 0.021 $). Since the Zipf distribution of the number of views is known, we can find the relationship between the hit-ratio and the number of cached videos. This relationship is plotted in Fig. 7. It shows that, to achieve a 90\% hit-ratio, the edge server would cache only 800 Douyin videos. As shown in Table 2, the average video size is about 2 MB. Therefore, caching 800 short videos only requires 1.56 GB of storage (about 19.53 GB if we use maximum size),
which will only be a small fraction of storage for an edge server.

\section{Conclusion and Future work}
We studied latest short videos from Douyin in terms three basic fundamental features and four popularity metrics. 
First, we studied how Douyin works and found out that it is a decentralized video social media based on massive short videos and a strong recommendation mechanism. 
Second, we studied the characteristics of Douyin videos and found out that the distributions of the bitrate and the size of videos closely follow Weibull distribution. 
Third, we analyzed distributions of different popularity metrics and discovered that the popularity metrics of the most popular Douyin videos obey Zipf distribution, but rest of the videos do not, which is the case with traditional videos. 
We also analyzed the relationships of key popularity metrics and figured out that four of them are not highly correlated, except the relationship between the number of views and the number of likes. 

Based on this work, we believe that our work provides an initial foundation for the design of future advanced caching systems for short video platforms. There are several possible directions for future work, that can be followed based on our work.
For instance, using deep reinforcement learning, we would like to design a proactive and effective caching system that can reduce the short video edge server load.

\bibliographystyle{ACM-Reference-Format}
\bibliography{reference}

\end{document}